\documentclass[pra,twocolumn,english,superscriptaddress,preprintnumbers,amsmath,amssymb]{revtex4}
\usepackage{graphicx}
\usepackage{wasysym}

\makeatletter
\@ifundefined{textcolor}{}
{%
 \definecolor{BLACK}{gray}{0}
 \definecolor{WHITE}{gray}{1}
 \definecolor{RED}{rgb}{1,0,0}
 \definecolor{GREEN}{rgb}{0,1,0}
 \definecolor{BLUE}{rgb}{0,0,1}
 \definecolor{CYAN}{cmyk}{1,0,0,0}
 \definecolor{MAGENTA}{cmyk}{0,1,0,0}
 \definecolor{YELLOW}{cmyk}{0,0,1,0}
}

\usepackage{amsfonts}
\usepackage{lipsum}
\usepackage{float}
\usepackage{placeins}

\DeclareMathOperator{\trace}{\mathrm{Tr}}
\DeclareMathOperator{\td}{\mathcal{I}}

\newcommand{\ket}[1]{\left|#1\right>}
\newcommand{\bra}[1]{\left<#1\right|}

\DeclareMathOperator{\nn}{\mathcal{N}(\Phi)}
\DeclareMathOperator{\nnm}{\mathcal{N}_{\mathrm{max}}(\mathit{N})}

\DeclareMathOperator{\gu}{\Gamma_{\uparrow}}
\DeclareMathOperator{\gd}{\Gamma_{\downarrow}}

\makeatother

\usepackage{babel}
\begin{document}

\title{Quantifying non-Markovianity due to driving and a finite-size environment in an open quantum system}

\author{Rui Sampaio}
\email{rui.ferreirasampaio@aalto.fi}

\selectlanguage{english}%
\affiliation{COMP Center of Excellence, Department of Applied Physics, Aalto University,
P.O. Box 11000, FI-00076 Aalto, Finland}

\author{Samu Suomela}
\affiliation{COMP Center of Excellence, Department of Applied Physics, Aalto University,
P.O. Box 11000, FI-00076 Aalto, Finland}

\author{Rebecca Schmidt}
\affiliation{COMP Center of Excellence, Department of Applied Physics, Aalto University,
P.O. Box 11000, FI-00076 Aalto, Finland}
\affiliation{Turku Centre for Quantum Physics, Department of Physics and Astronomy,
University of Turku, FIN-20014 Turku, Finland}
\affiliation{Center for Quantum Engineering, Department of Applied Physics, Aalto
University School of Science, P.O. Box 11000, FIN-00076 Aalto, Finland}

\author{Tapio Ala-Nissila}
\affiliation{COMP Center of Excellence, Department of Applied Physics, Aalto University,
P.O. Box 11000, FI-00076 Aalto, Finland}
\affiliation{Department of Physics, Box 1843, Brown University, Providence, Rhode Island 02912-1843, U.S.A.}

\begin{abstract}
We study non-Markovian effects present in a driven qubit coupled to a finite environment using a recently proposed model developed in the context of calorimetric measurements of open quantum systems. To quantify the degree of non-Markovianity we use the Breuer-Laine-Piilo (BLP) measure [Phys. Rev. Lett. \textbf{103}, 210401 (2009)]. We show that information backflow only occurs in the case of driving in which case we investigate the dependence of memory effects on the environment size, driving amplitude and coupling to the environment. We show that the degree of non-Markovianity strongly depends on the ratio between the driving amplitude and the coupling strength. We also show that the degree of non-Markovianity does not decrease monotonically as a function of the environment size.

\end{abstract}
\maketitle

\section{Introduction}
While the Markovian description is often a very good approximation of open system's evolution in quantum optics \cite{Breuer2002}, the underlying assumptions, such as a weak system-environment coupling  and a memoryless i.e. infinite and not too cold environment can be easily violated in condensed matter systems \cite{PhysRevE.94.022123,Pekola:2015aa, RS2015}. For such systems a detailed understanding of the memory effects is desirable. 
The study of non-Markovianity in open quantum systems has become a subject of broad interest in recent years due to the rapid development in  quantum information \cite{rivas2014quantum, breuer2016colloquium} and in quantum thermodynamics \cite{Esposito2009,Campisi20111,PhysRevE.87.032113,Pekola:2013, Suomela:2016,Suomela:2016aa,Pekola:2016, katz2016quantum, bylicka2016thermodynamic,strasberg2016nonequilibrium,
newman2016performance,kato2016quantum,whitney2016non,PhysRevA.94.010101,uzdin2016zeno}. 
There are a great variety of non-Markovianity measures for open quantum systems proposed in the literature \cite{prl103/210401,laine2010measure, rivas2010entanglement,lu2010quantum, vasile2011quantifying, luo2012quantifying,  lorenzo2013geometrical, bylicka2014non, addis2014comparative}, which are not identical. Some of them, such as the RHP (Rivas-Huelga-Plenio \cite{rivas2010entanglement}) measure are based on the non-divisibility of the dynamical map, while other quantifiers focus on the information backflow, such as the BLP (Breuer-Laine-Piilo \cite{prl103/210401}) which is based on the evolution of the distinguishability of quantum states. The BLP measure is commonly used, as its operational definition allows for a clear physical interpretation of information backflow.

Non-Markovian dynamics is commonly associated with a strong coupling between the system and the environment or parts of the environment \cite{PhysRevA.79.042302}.
However, the dynamics can become non-Markovian even in the weak coupling regime if the environment is either very cold \cite{PhysRevA.94.010101} or finite \cite{PhysRevA.83.032103,PhysRevLett.107.080404,jeske2012dual,uzdin2016zeno}. This is especially important in stochastic thermodynamics of open quantum systems, where non-Markovianity plays a crucial role in the detection of heat and work. In stochastic thermodynamics, direct detection of work and heat requires that the heat exchange between the system and the environment leaves detectable traces to the environment. One proposed measurement scheme to do this is the calorimetric detection of the immediate environment \cite{Pekola:2013,Viisanen2015}.

In the calorimetric measurement scheme, heat exchange between the system and the environment is obtained by monitoring the environment's energy or effective temperature. However, in order to witness the changes in the environment's state, the environment has to be finite in contrast to an infinitely large or memoryless environment required for justifying the Markovian approach. For this reason, the calorimetric setup cannot be modelled with the standard Lindblad master equation. 

In recent articles \cite{Suomela:2016,Suomela:2016aa,Pekola:2016}, a modified Lindblad-like equation was introduced that takes into account the finite-size of the environment. This model is suitable to describe the calorimetric measurement. The corresponding finite environment master equation (FEME) \cite{Suomela:2016aa} and its stochastic unraveling \cite{Suomela:2016} lead to stochastic trajectories of the system state that depend on their own history via the state of the environment. Thus, from the point of view of the system's degrees of freedom, the trajectories are non-Markovian \footnote{This non-Markovianity clearly differs from the non-Markovian trajectories of the Non-Markovian quantum jump model \cite{Piilo:2008aa}, where the evolution of a trajectory does not depend on its own history but only on the collective history of all the trajectories.}. 

In this article, we study the degree of non-Markovianity induced by the finite-size of the environment. We analyze the non-Markovianity of FEME by using the BLP measure.
We focus on a driven qubit coupled to an environment consisting of two-level systems with the same energy gap. We assume that the environmental degrees of freedom decohere faster than any other time scale such that we can use FEME. With the BLP  measure, we quantify the level of non-Markovianity for different environment sizes, driving strengths and system-environment coupling strengths. We show that FEME leads to non-Markovian dynamics according to the BLP measure in the presence of driving. We additionally show that the degree of non-Markovianity does not always decrease monotonically as a function of the environment size.

\section{Model}

\begin{figure}
\centering \includegraphics[width=\linewidth]{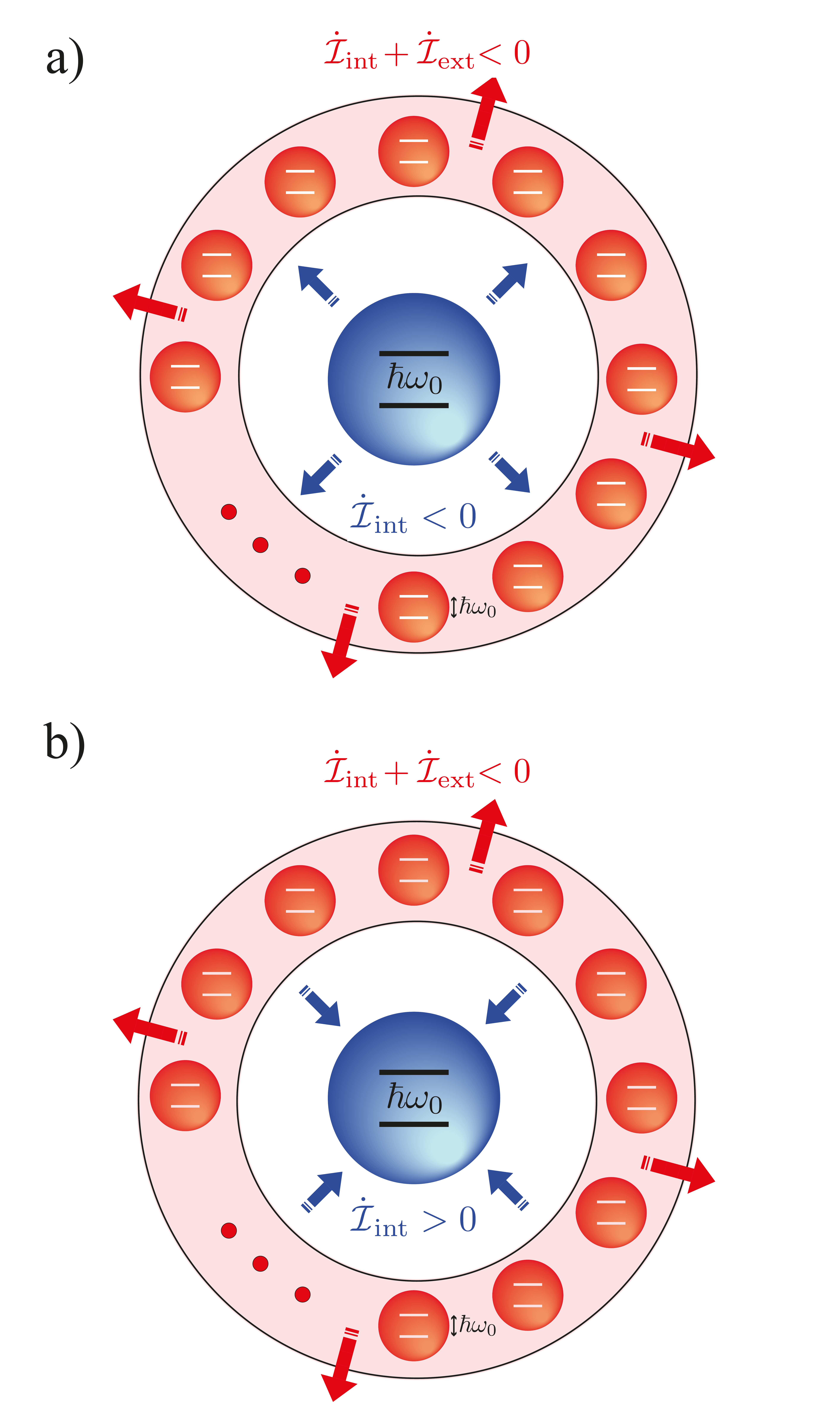} \caption{(Color online) Schematic illustration of a qubit interacting with the calorimeter. The calorimeter is composed of a finite number of two-level systems, resonant with the qubit. The arrows illustrate the direction of the flow of information. $ \td_\text{int} $ represents the information in the qubit and $ \td_\text{ext} $ represents the information in the calorimeter (see text for details). a) Markovian case where there is constant loss of information by the qubit and by the whole system. b) Non-Markovian case where there is back-flow of information into the qubit but the system, as a whole, still behaves as Markovian.}
\label{fig:model} 
\end{figure}

We focus on a driven qubit (two level system) with Hamiltonian $H_\text{q}(t)=\hbar\omega_{0}a^{\dagger}a+\lambda(t)(a^{\dagger}+a)$,
where $a$ ($a^{\dagger}$) is the annihilation (creation) operator,
$\hbar\omega_{0}$ is the qubit energy gap and $\lambda(t)$ is a
real valued function representing the driving protocol. The qubit interacts with
a finite environment comprised of $N$ two-level systems with an energy gap identical to that of the qubit (see Fig. \ref{fig:model}). Let us refer to the environment as the calorimeter from here on. We assume the qubit to be weakly coupled to the calorimeter by $V=\sum_{k=1}^{N}( \kappa_k  {a}^\dagger {d}_k  + \kappa_k^* {a} {d}_k^\dagger$), where $\kappa_k$ is the coupling strength and $ d_k$ and $ d^\dagger_k$ are the calorimeter's annihilation and creation operators of the $k^\text{th}$ two-level system. The calorimeter Hamiltonian is given by $  H_\text{c} =\sum_{k=1}^{N} \hbar \omega_0 d_k^\dagger d_k$. 

Similar to the Refs. \cite{Suomela:2016,Suomela:2016aa,Pekola:2016}, we assume the calorimeter to decohere quickly into a microcanonical ensemble such that the total density matrix of the qubit-calorimeter composite  can be written as
\begin{equation}
\rho_\text{QC}(t)=\sum_{n=0}^N \sigma(n,t) \otimes \sigma_\text{c}(E_n),
\label{eq:sigma_QC(t)}
\end{equation}
where $n$ denotes the number of excited two-level systems in the calorimeter, $E_n = n \hbar \omega_0$ denotes the corresponding calorimeter energy, $\sigma_c(E_n)$ is the microcanonical ensemble of calorimeter microstates corresponding to the energy $E_n$, \textit{i.e.}, $\sigma_c(E_n) =[1/N(E_n)] \sum_{k}  \ket{\Psi_k} \bra{\Psi_k} \delta_{\epsilon_k,E_n}$, where $\epsilon_k$ is the energy of microstate $\ket{\Psi_k}$ and  $N(E_n)$ is the number of microstates with energy $E_n$. It is important to note that $\sigma(n,t)$ cannot be interpreted as a reduced density matrix. Instead, its trace represents the probability $p_{n}(t)$ of finding the calorimeter with a given energy $E_n$ as a function of time. The reduced density matrix is obtained by tracing over all the calorimeter ($\text{c}$) degrees of freedom: 
\begin{equation}
\begin{split}
\sigma(t) &=\trace_\text{c} \left\lbrace \rho_\text{QC}(t) \right\rbrace =\sum_{n=0}^{N}\sigma(n,t).
\end{split}
\label{eq:sigma(t)}
\end{equation}
Treating the drive and qubit-calorimeter coupling perturbatively, it can be shown \cite{Suomela:2016aa} that the evolution of $\sigma(n,t)$ obeys a finite environment master equation (FEME):
\begin{equation}
\begin{split}
\dot{\sigma}(n,t) =& \frac{i}{\hbar}\left[\sigma(n,t),H_\text{q}(t)\right] - \frac{\Gamma_{\uparrow}(n)}{2}\left\{ \sigma(n,t),aa^{\dagger}\right\} \\
 &+\Gamma_{\downarrow}(n-1)a\sigma(n-1,t)a^{\dagger} \\
 &+\Gamma_{\uparrow}(n+1)a^{\dagger} \sigma(n+1,t) a\\
 &-\frac{\Gamma_{\downarrow}(n)}{2}\left\{ \sigma(n,t), a^{\dagger}a\right\},
\end{split}
\label{eq:sigma_n-dot}
\end{equation}
where $\Gamma_{\uparrow}(n)$ and $\Gamma_{\downarrow}(n)$ are the
transition rates associated with the Lindblad operators $a^{\dagger}$
and $a$, respectively. The transition rates depend on the calorimeter
energy $E_{n}$ and are expressed as 
\begin{align}
\Gamma_{\downarrow}(n) & =g(1-n/N);\label{eq:gamma_d}\\
\Gamma_{\uparrow}(n) & =gn/N,\label{eq:gamma_u}
\end{align}
where $g$ is a tunable parameter representing the coupling strength
between the system and the calorimeter. Because these transition rates
depend on the energy of the calorimeter, the evolution of the qubit depends on its history.

In general, Eq. (\ref{eq:sigma_n-dot}) produces a divisible map only for the set of $\sigma(n,t)$, but not for the reduced density matrix of the qubit. Thus, the dynamics of the latter can be non-Markovian, according to the divisibility definition of Markovianity  \cite{rivas2010entanglement,rivas2014quantum}.

In order to quantitatively study non-Markovianity and the degree of it generated by Eq. (\ref{eq:sigma_n-dot}), we use the BLP measure that focuses on the distinguishability of quantum states. One reason to choose the BLP measure is that if the dynamics is non-Markovian according to the BLP measure, then the dynamics is non-Markovian also according to other measures such as the non-divisibility  and the semigroup properties \cite{rivas2014quantum}. Formally, the BLP measure
$\mathcal{N}(\Phi)$ is given by \cite{Breuer:2009aa,breuer2016colloquium}
\begin{equation}
\mathcal{N}(\Phi)=\underset{\sigma^{1,2}}{\mathrm{max}}\int_{\dot{\mathcal{I}}(t)>0}\mathrm{d}t\dot{\mathcal{I}}(t),\label{eq:nn}
\end{equation}
where $\mathcal{I}(t)\equiv\mathcal{I}(\Phi_{t}\sigma^{1},\Phi_{t}\sigma^{2})=
\|\Phi_{t}(\sigma^{1}-\sigma^{2})\|/2$ is the trace distance between $\Phi_{t}\sigma^{1,2}$ as a function of time, and $\Phi_{t}$ is a linear map of the reduced density matrix associated with the formal solution of the set of Eq. (\ref{eq:sigma_n-dot}), \textit{i.e.}, $\sigma(t)=\Phi_{t}\sigma(0)$. The integral is taken over all the regions where the trace
distance rate $\dot{\mathcal{I}}(t)$ is positive and the maximization
is performed over all possible pairs of initial states for the reduced density matrices.
For the particular case of the qubit, it has been shown
that the optimal pair must be pure orthogonal states \cite{Wismann:2012aa}. Therefore, the
initial condition $\sigma^{1}-\sigma^{2}\equiv\tilde{\sigma}_{0}$
can be written as 
\begin{equation}
\tilde{\sigma}_{0}=\left(\begin{array}{cc}
\cos\theta & e^{i\phi}\sin\theta\\
e^{-i\phi}\sin\theta & -\cos\theta
\end{array}\right),\label{eq:sigma(0)}
\end{equation}
where $\theta$ are $\phi$ are the usual Bloch sphere angles. With
these simplifications, the maximization is taken over $\theta\in[0,\pi[$
and $\phi\in[0,\pi[$. In all of the following results, the calorimeter
starts from canonical equilibrium at inverse temperature $\beta$ such
that 
\begin{equation}
\tilde{\sigma}(n,0)=p_{n}(0)\tilde{\sigma}_{0},\label{eq:sigma_n(0)}
\end{equation}
with $p_{n}(0)=\binom{N}{n}\exp[-\beta(E_{n}-F)]$, where $F$ is the free energy.
Furthermore, we use a sinusoidal driving protocol $\lambda(t)=\lambda_{0}\sin(\omega_{0}t)$
resonant with the qubit (see Appendix \ref{sec:max} for details on the evaluation of $\nn$).

\section{Results}

\subsection{Trace distance behavior}

\begin{figure}
\begin{center}
\includegraphics[width=0.9\linewidth]{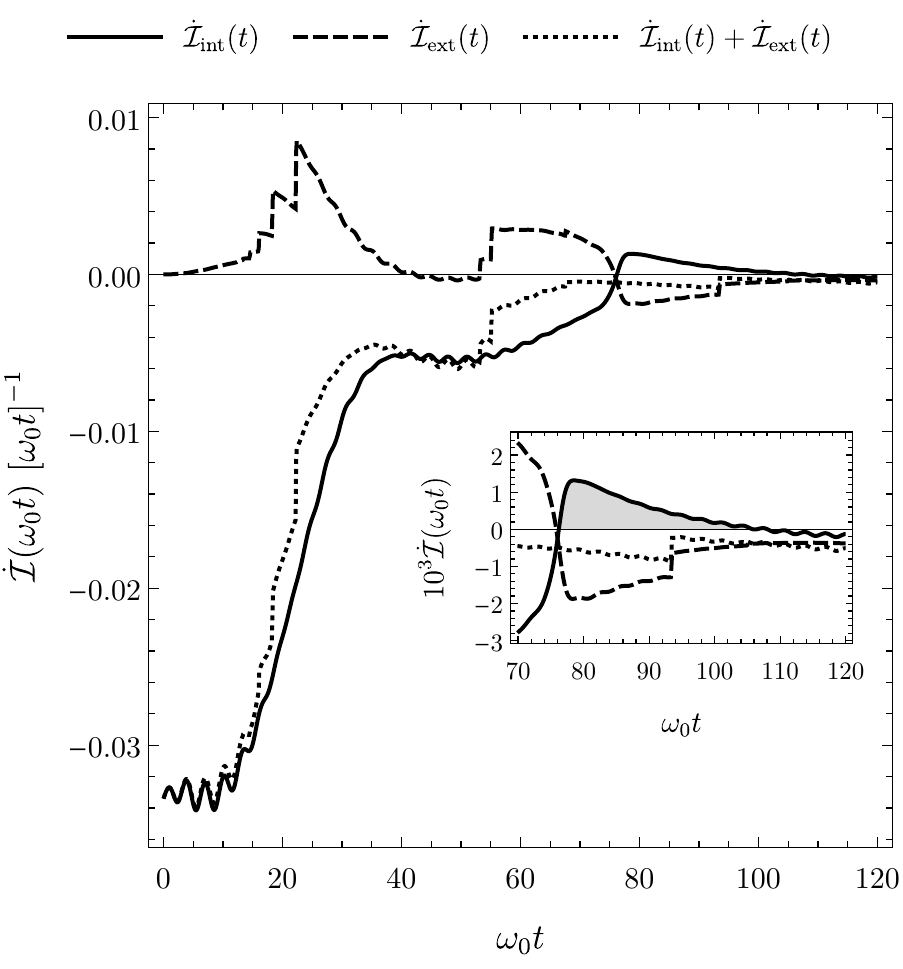}
\end{center}
\caption{Rate of change of the information in the qubit, $ \dot{\td}_\text{int} $, calorimeter, $ \dot{\td}_\text{ext} $, and total information, $ \dot{\td}_\text{int}+ \dot{\td}_\text{ext} $, as a function of time. The inset highlights the region of the evolution that contributes the BLP measure in Eq. (\ref{eq:nn}). The parameters used are $ \beta\hbar\omega_0 = 2 , \lambda_0/\hbar\omega_0 = 0.08 , g/\hbar\omega_0 = 0.066, N = 20, \theta = 1.69 $ and $ \phi = 0  $. }
\label{fig:I(t)}
\end{figure}

We start by discussing the main characteristics of $\dot{\td}(t)$
for the system studied. Figure \ref{fig:I(t)} shows a typical example for
a particular set of parameters (see the figure caption for details). For
consistency, we employ the notation in Ref. \cite{breuer2016colloquium}, denoting
$\mathcal{I}_{\text{int}}\equiv\mathcal{I}$ and $\mathcal{I}_{\text{ext}}=D(\rho_{\text{QC}}^{1}(t),\rho_{\text{QC}}^{2}(t))-\mathcal{I}_{\text{int}}$,
where $D(\rho_{\text{QC}}^{1}(t),\rho_{\text{QC}}^{2}(t))$ is the
trace distance between $\rho_\text{QC}^{1}$ and $\rho_\text{QC}^{2}$. 
 The notation emphasizes the connection to the information theoretic interpretation of the BLP measure. $\td_{\text{int}}(t)$ represents the distinguishability of the qubit reduced state while $\td_{\text{ext}}(t)$ represents the distinguishability of the total system states minus the distinguishability of the qubit states. For brevity, we shall refer to $\td_{\text{int}}(t)$ and $\td_{\text{ext}}(t)$ as the information in the qubit and calorimeter, respectively.

For a closed system one expects the total information $\td_{\text{int}}(t)+\td_{\text{ext}}(t)$ to be conserved. However, it is clear from Fig. \ref{fig:I(t)} that is not the case here. In fact, there is constant loss of information, given that $\dot{\td}_{\text{int}}(t)+\dot{\td}_{\text{ext}}(t)$ is always negative. This is due to the total system's non-unitary evolution produced by Eq. \eqref{eq:sigma_n-dot}, where we have assumed the calorimeter to be decohered. 
Other features that draw immediate attention are the discontinuities of $\dot{\td}_{\text{ext}}(t)$. They arise from the fact that the eigenvalues of the difference between $\rho_\text{QC}^{1}$ and $\rho_\text{QC}^{2}$ can cross between positive and negative values. By definition, the trace distance is not differentiable at these points, which leads to the discontinuities in Fig. \ref{fig:I(t)}. For the qubit trace distance, discontinuities are not present because one of the eigenvalues of $\Phi_{t}\tilde{\sigma}_{0}$ is always positive and the other always negative.

Contrary to other systems where non-Markovianity has been investigated \cite{prl103/210401, breuer2016colloquium,PhysRevA.94.010101}, the trace distance rate starts from negative values. In those works, the  time-dependent transition rates are continuous at time $t=0$ and start from zero. However, in our case, the transition rates [Eqs. (\ref{eq:gamma_d}) and (\ref{eq:gamma_u})] are positive and time-independent which leads to discontinuous transition rates at time $t=0$ \footnote{The same behavior occurs using the Lindblad equation with time-independent transition rates.}. 

It is important to note that non-Markovianity in this system is induced by the driving. If no driving is present, the set of Eqs. (\ref{eq:sigma_n-dot}) can be solved exactly and the trace distance is given by,
\begin{equation}\label{key}
\td(t) = \sqrt{f(t)^2\cos^2\theta + g(t)^2\sin^2\theta},
\end{equation}
where $ f(t) = (N\exp[-g(1+1/N)t] + 1)/(N+1) $ and $ g(t) = \exp(-gt/2) $ and we have explicitly assumed the initial condition in Eq. (\ref{eq:sigma_n(0)}). Since $ f(t) $ and $ g(t) $ are monotonically decreasing functions of $ t $, $ \nn = 0 $. Thus, the dynamics are Markovian according to the BLP measure in the case of no driving. However, it should be noted that even in this case, Eq. (\ref{eq:sigma_n(0)}) generally produces dynamics that are non-divisible for the reduced density matrix and thus the dynamics become non-Markovian according to measures based on the non-divisibility, such as the RHP measure. 

The inset in Fig. \ref{fig:I(t)} highlights the part of the evolution that contributes to the BLP measure. Pairs that maximize the BLP measure present sharp transitions from negative to positive values of $\dot{\td}(t)$. At this point information starts flowing back into the qubit. 

The BLP measure considers only maximization over initial state pairs. From the experimental point of view it is interesting to consider how the parameters of the system affect $ \nn $. For our system, there are four such parameters \textendash{} the calorimeter size $N$, the coupling strength between the qubit and the calorimeter $g$, the drive strength $\lambda_{0}$, and the inverse temperature $ \beta $. In the next section we examine how $\nn$ depends on the first three parameters. We limit ourselves to the case of low temperatures such that $ \beta\hbar\omega_0 > 1  $ since the calorimetric detection works in this regime. In all of the following results, temperature is fixed according to $ \beta\hbar\omega_0 = 2 $.

\subsection{Influence of the system parameters on $\nn$}

\begin{figure}[t!]
\centering \includegraphics[width=0.9\linewidth]{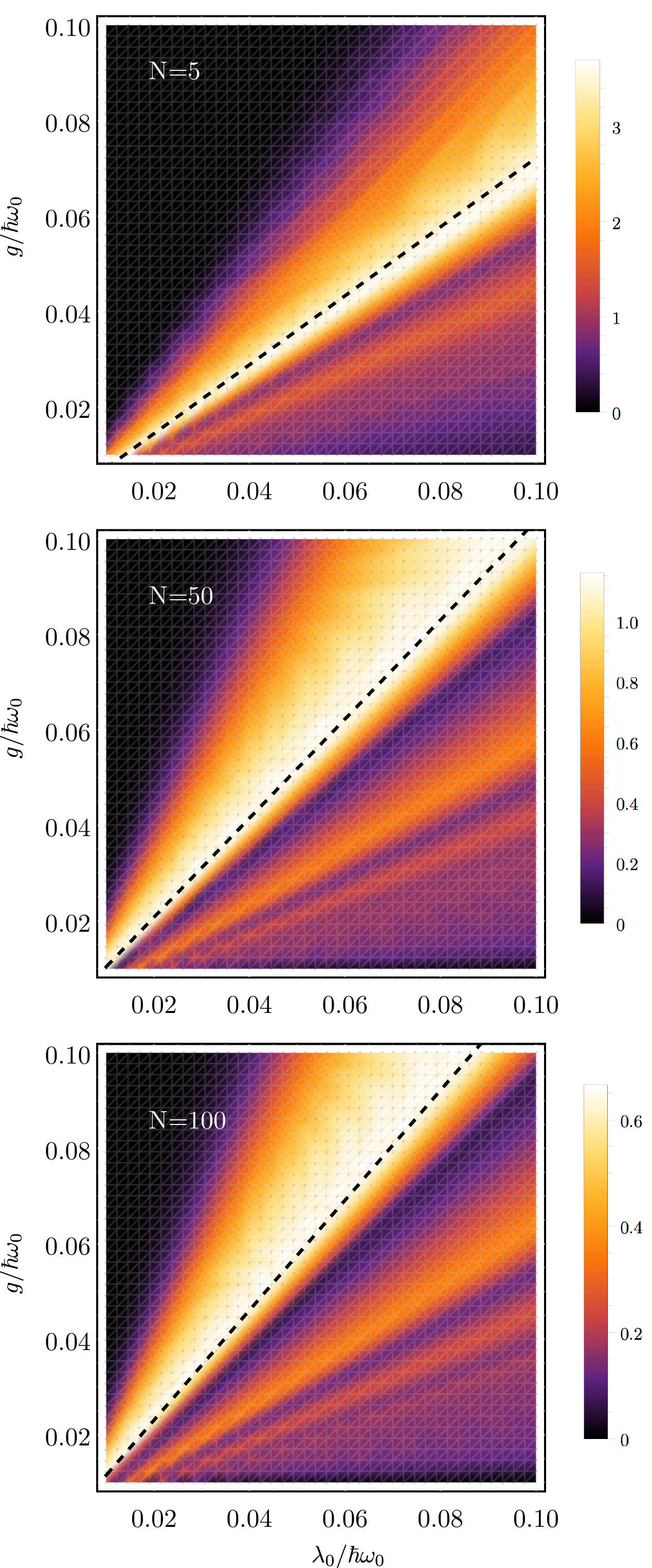}
\caption{(Color online) BLP measure ($10^{2}\mathcal{N}(\Phi)$, color scale) as a function
of the drive strength $\lambda_{0}$ and the coupling constant $g$
in a grid of 40$\times$40 points for 5 (top),  50 (middle) and 100
(bottom) two-level systems in the calorimeter. The dashed line indicates the line
of constant maximum $\nnm$, given by $\lambda_{0}/g=a_{N}$. }
\label{fig:parameter_sweep} 
\end{figure}

To show how the degree of non-Markovianity depends on the system parameters, we plot $ \nn $ in Fig. \ref{fig:parameter_sweep} as a function of $\lambda_{0}$ and $g$, for three different calorimeter sizes, $N=5$ (top), $50$ (middle) and $100$ (bottom). A non-Markovian structure emerges characterized by "rays" of constant $\nn$. An interesting aspect is the oscillatory nature of this structure. It shows that one cannot deduce that a smaller calorimeter will necessarily always induce an higher degree of non-Markovianity as compared to a larger calorimeter. In fact, for fixed $\lambda_{0}$ and $g$ we can go from Markovian to non-Markovian behavior by increasing the calorimeter size. On the other hand, at a fixed calorimeter size, one can choose between Markovian or non-Markovian behavior by tuning the coupling to the environment or the driving strength. 

The figure of merit for a particular value of $ N $ is the peak value of $ \nn $, which we label as $ \nnm $.   
Depending on the specific application one may be interested in asking, for example, how does $\nn$ change for fixed system parameters, how does $\nnm$ change within a certain range of parameters, or how does one assure that non-Markovian effects are mitigated. Here, we will focus on how $ \nnm $ changes as a function of the calorimeter size.

\begin{figure}
\centering\includegraphics[width=0.9\linewidth]{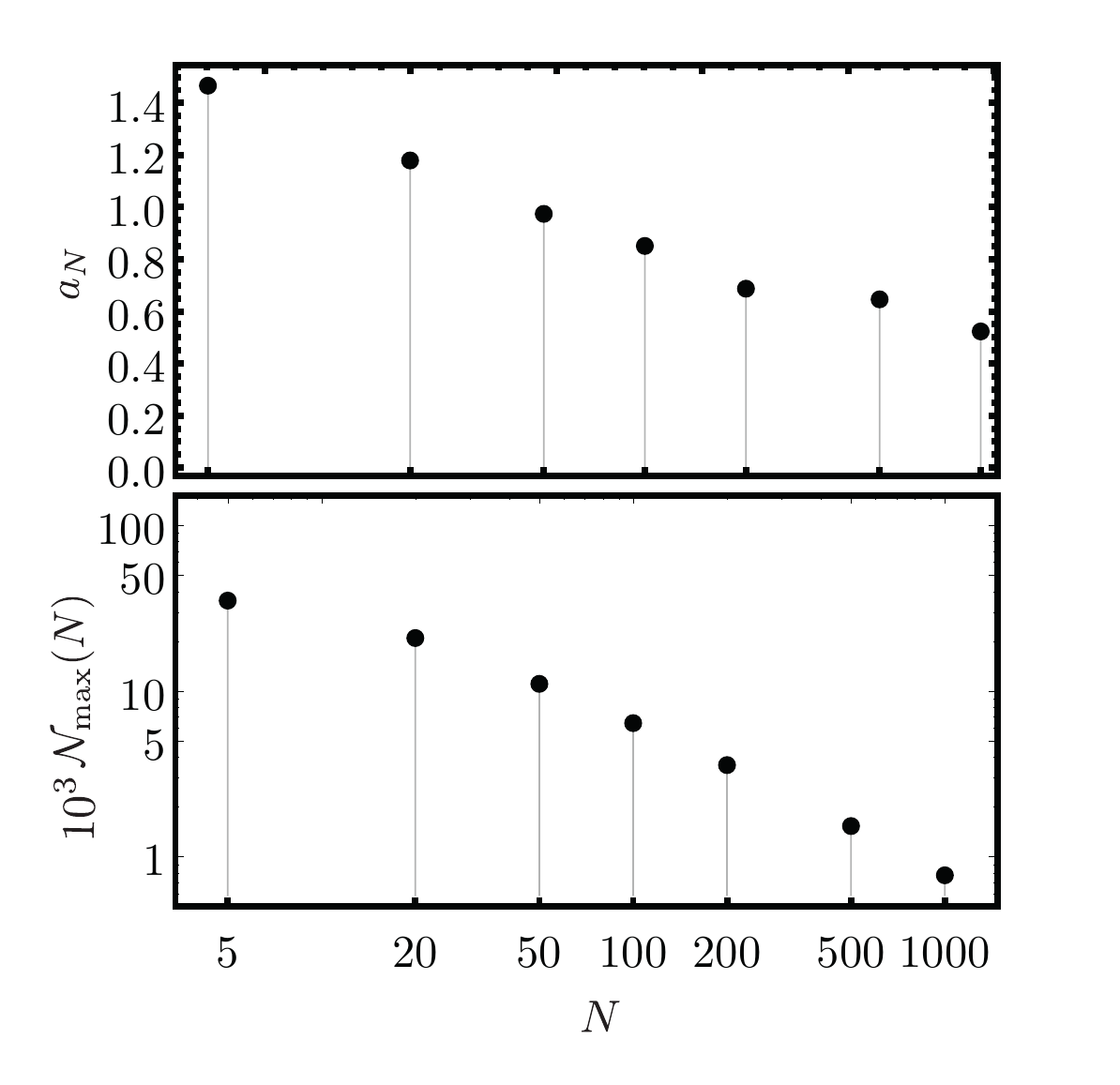} \caption{A log-linear plot of the ratio $a_{N}$ and log-log plot of $\nnm$ as a function of the calorimeter size $N$.}
\label{fig:nn_size} 
\end{figure}

The peak value $ \nnm $ falls in the straight line $\lambda_{0}/g=a_{N} $ (cf. Fig. \ref{fig:parameter_sweep}). From Fig \ref{fig:nn_size}, we see that  $\nnm$ decreases as the calorimeter size increases and seems to converge to $\nnm\rightarrow0$  as $N\rightarrow\infty$. This is expected as the relative energy fluctuations around the calorimeter average energy become zero for $N\rightarrow\infty$. As the driving time and amplitude are finite, the transition rates become constant as the changes in the calorimeter energy due to the driving are negligible compared to the initial average energy for $N\rightarrow\infty$. Consequently, in this limit the dynamics of the reduced density matrix converge to a Lindblad master equation. We also notice that the value of $ \nnm $ is rather small for large sizes of the calorimeter. For $ N = 10^3 $, $ \nnm \sim 10^{-3} $. This means that the amount of information recovered is not significant. 

\begin{figure}
\centering \includegraphics[width=1\linewidth]{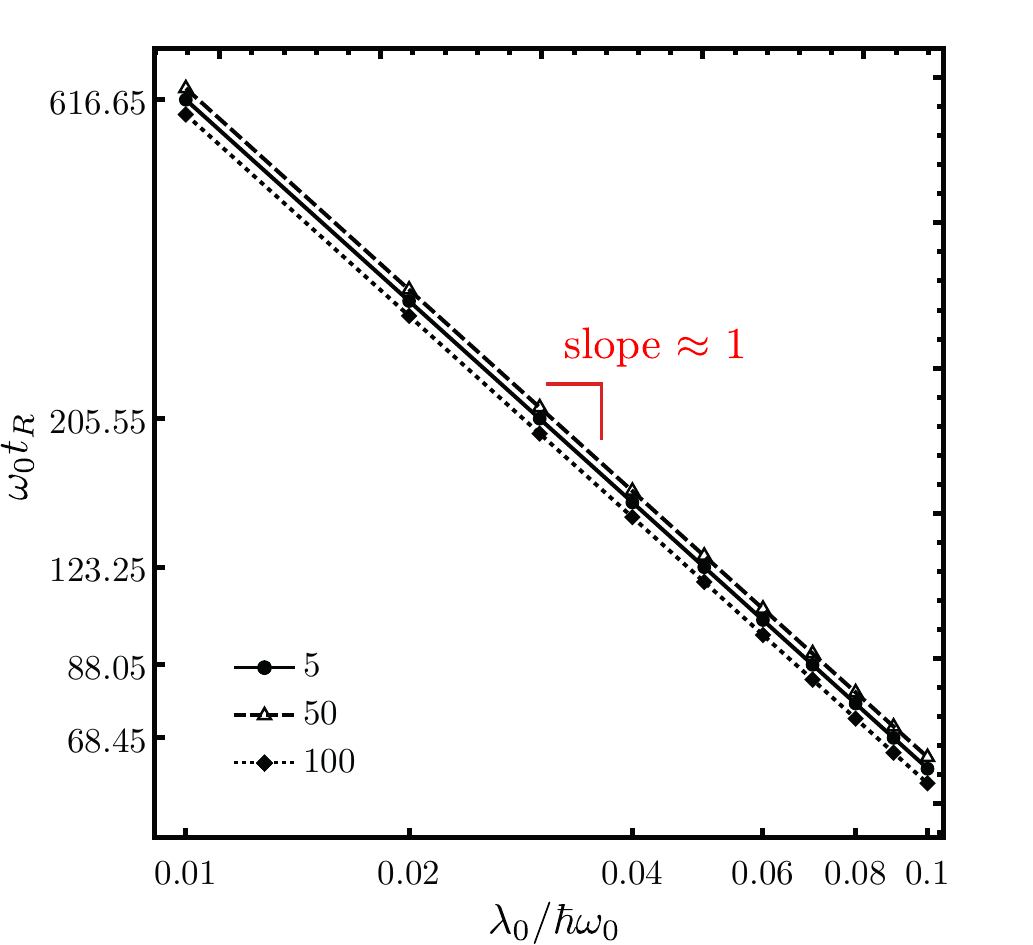} \caption{A log-log plot of $\omega_{0}t_{R}$ along the line of maximum $\nn$ ($\lambda_{0}=a_{N}g$) for the three different
calorimeter sizes $N=5$, $50$ and $100$. }
\label{fig:tr} 
\end{figure}

Finally, as highlighted in Fig. \ref{fig:I(t)}, $ \dot\td $ exhibits a sharp transition from negative to positive at a given time, denoted by $t_{R}$. From the information theoretic or reservoir engineering perspective it is useful to know the first point in time at which information flows back to the system. As in the case of $ \nn $, $ t_R $ varies with the parameter used. However, if we focus on the line of $ \nnm $, a clear pattern emerges. Figure \ref{fig:tr} shows a $\log-\log$ plot of $t_{R}$ along the line $\lambda_{0}/g=a_{N}$, as a function of $\lambda_{0}$, for the three calorimeter sizes $N=5,50$ and $100$. Remarkably, $t_{R}$ is almost independent of the calorimeter size and we can write 
\begin{equation}
t_{R}\propto\frac{\hbar}{\lambda_{0}}.\label{eq:tR}
\end{equation}
Note that this is not true if $\lambda_{0}$ and $g$ are held fixed
and the size $N$ is varied. In Fig. \ref{fig:tr} and Eq. (\ref{eq:tR}),
for a given $\lambda_{0}$ and $N$, the coupling $g$ is implicitly
given by $g=\lambda_{0}/a_{N}$. That is, the time at which information starts to flow back into the system can be easily tuned by adjusting the driving amplitude.  

\section{Summary and Conclusions}

In this article, we have studied the non-Markovianity of the recently proposed FEME, which includes finite-size effects of the environment. We focused on a driven qubit coupled to an environment consisting of two-level system with the same energy gap as the qubit. We have shown that the FEME produces non-Markovian dynamics according to the BLP measure in the presence of an external drive. This implies that the model produces non-Markovian dynamics also with other definitions of Markovianity based on  non-divisibility and semigroup properties \cite{rivas2014quantum}.

With the BLP measure, we have shown that the degree of non-Markovianity strongly depends on the ratio of the driving amplitude and the qubit-environment coupling strength as witnessed in Fig. \ref{fig:parameter_sweep}. Suprisingly, for a fixed value of the driving amplitude and the coupling strength, the degree of Non-Markovianity does not decrease monotonically as a function of the number of two-level systems in the environment. However, if maximized over all the possible values of the coupling strength and the drive amplitude, the degree of non-Markovianity tends to decrease exponentially towards zero as a function of the environment size. For this reason, the amount of re-coherence or information recovered is relatively small for very large environments.

We also investigated the behavior of the time ($t=t_R$) when the information backflow first occurs. We showed that when non-Markovianity is maximized over the coupling strength, $t_R$ decreases linearly as a function of the drive amplitude for all environment sizes. This means that the occurence of information backflow can be easily shifted by adjusting the driving field. That is, if non-Markovianity is harmful to the performance of the system, one can shift the occurrence of information backflow to time values larger than the operation time. On the other hand, if one wants to use the information backflow as a resource, one can shift it to occur at a time where its effect on the system performance is the largest. 

\acknowledgements{
We thank Jyrki Piilo for fruitful discussions. We gratefully acknowledge financial support by the Academy of Finland through project number 287750, the Centers of Excellence Programme (2015-2017) under project numbers 251748 and 284621, the Centre of Quantum Engineering at Aalto University School of Science. S.S. acknowledges financial support from the V\"{a}is\"{a}l\"{a} foundation. The numerical calculations were performed using computer resources of the Aalto University School of Science "Science-IT" project.
}

\FloatBarrier 
\bibliography{finite-cal-nonMarkov}

\FloatBarrier \onecolumngrid

\appendix

\section{Maximization in the BLP measure}

\label{sec:max}

To perform the maximization of $\nn$ in Eq. (\ref{eq:nn}) in the
main text, we have to solve the system of equations (\ref{eq:sigma_n-dot}).
To this end, we work in the interaction picture with respect to 
$ H_0 \equiv \hbar\omega_0 a^\dagger a $.  Then, one has to solve
the evolution dictated by the $4(N+1)$ equations, 
\begin{equation}
\begin{aligned}
\dot{\sigma}_{00}^{n}(t) & = iv^*\sigma_{01}^{n}(t)-iv\sigma_{10}^{n}(t)
-\Gamma_{\uparrow}(n)\sigma_{00}^{n}(t)
+\Gamma_{\downarrow}(n-1)\sigma_{11}^{n-1}(t);  \\
\dot{\sigma}_{11}^{n}(t) & =iv\sigma_{10}^{n}(t)-iv^*\sigma_{01}^{n}(t) 
- \Gamma_{\downarrow}(n)\sigma_{11}^{n}(t)
+\Gamma_{\uparrow}(n+1)\sigma_{00}^{n+1}(t); \\
\dot{\sigma}_{01}^{n}(t) & =iv(\sigma_{00}^{n}(t)-\sigma_{11}^{n}(t))
-\frac{\Gamma_{\Sigma}}{2}\sigma_{01}^{n}(t); \\
\dot{\sigma}_{10}^{n}(t) & =iv^*(\sigma_{11}^{n}(t)-\sigma_{00}^{n}(t))
-\frac{\Gamma_{\Sigma}}{2}\sigma_{10}^{n}(t),
\end{aligned}
\label{eq:dot-sigma_apdx}
\end{equation}
where $\Gamma_{\Sigma}=\gu(n)+\gd(n)=g$, $\sigma_{ij}^{n}(t)=\left\langle i|\sigma(n,t)|j\right\rangle $ and $ v = \lambda_0 \exp[-i\omega_0t]\sin(\omega_0t) $.
From Eq. (\ref{eq:sigma(t)}), the trace distance rate $\dot{\mathcal{I}}(t)$
is given by 
\begin{equation}
\dot{\mathcal{I}}(t)=\frac{1}{\mathcal{I}(t)}\left(\sigma_{00}(t)\frac{\rm{d}\sigma_{00}(t)}{\rm{d}t}+|\sigma_{01}(t)|\frac{\rm{d}|\sigma_{01}(t)|}{\rm{d}t}\right),
\end{equation}
 where $\sigma_{ij}(t)=\sum_{n}\sigma_{ij}^{n}(t)$. This is then
integrated over all regions where $\dot{\mathcal{I}}(t) >0$ for a particular
initial condition, given by Eqs. (\ref{eq:sigma(0)}) and (\ref{eq:sigma_n(0)})
in the main text. In practice, we have to truncate the solution after
some time $\tau$. We have used $\omega_{0} \tau=1000\pi$, after which all
the solutions show exponential decay. Finally, for
given parameters $\lambda_{0}$ and $g$ we sample half of the Bloch
sphere in steps of $0.08 \mathrm{rad}$ in both $\theta$ and $\phi$. 
The maximum always appears for $\phi=0$ or $\phi=\pi$. The only exception is
when the maximum is along $\theta=0$ or $\theta=\pi$ for which the
solution is independent of $\phi$. This suggests that the optimal
pair lines up with the plane formed by the driving term eigenvectors and the
undriven qubit Hamiltonian eigenvectors. To further
test this, we change the coupling of the driving to be proportional
to the Pauli $\sigma_{y}$ matrix, resulting in the optimal pair to
be in the $yz$ plane (results not shown). Numerical evidence shows that
all solutions share this symmetry and we therefore assume in all the calculations
that the optimal pair is in the $xz$ plane of the Bloch sphere by
setting $\phi=0$. 

\end{document}